# Continuous use of ERP-based BCIs with different visual angles in ALS patients

**Jing Jin**[1] *, **Brendan Z. Allison**[3], **Yu Zhang**[1], **Yan Chen**[2]*, **Sijie Zhou**[1], **Yi Dong**[2], **Xingyu Wang**[1] **and Andrzej Chchocki** [4, 5, 6]

[1] Key Laboratory of Advanced Control and Optimization for Chemical Processes, Ministry of Education, East China University of Science and Technology, Shanghai, China

[2] Department of Neurology, Huashan hospital, Fudan University, Shanghai, P.R. China

[3] Department of Cognitive Science, University of California at San Diego, La Jolla, California, USA

[4] Laboratory for Advanced Brain Signal Processing, Brain Science Institute, RIKEN, Wako-shi, Japan，

[5] Systems Research Institute PAS, Warsaw

[6] Nicolaus Copernicus University (UMK), Torun, Poland

*Corresponding author: E-mail: jinjingat@gmail.com, chhyann@163.com

**Abstract**

*Objective*: Amyotrophic lateral sclerosis (ALS) is a rare disease, but is also one of the most common motor neuron diseases, and people of all races and ethnic backgrounds are affected. There is currently no cure. Brain computer interfaces (BCIs) can establish a communication channel directly between the brain and an external device by recognizing brain activities that reflect user intent. Therefore, this technology could help ALS patients in promoting functional independence through BCI-based speller systems and motor assistive devices. *Methods*: In this paper, two kinds of ERP-based speller systems were tested on 18 ALS patients to: (1) assess performance when they spelled 42 characters online continuously, without a break; and (2) to compare performance between a matrix-based speller paradigm (MS-P, mean visual angle 6˚) and a new speller paradigm that used a larger visual angle called the large visual angle speller paradigm (LS-P, mean visual angle 8˚). *Results*: Although results showed that there were no significant differences between the two paradigms in accuracy trend over continuous use ($p>0.05$), the fatigue during the LS-P condition was significantly lower than that of MS-P ($p<0.05$). Results also showed that continuous use slightly reduced the performance of this ERP-based BCI. Conclusion: 15 subjects obtained higher than 80% feedback accuracy (online output accuracy) and 9 subjects obtained higher than 90% feedback accuracy in one of the two paradigms, thus validating the BCI approaches in this study. *Significance*: Most ALS subjects in this study could spell effectively after continuous use of an ERP-based BCI. The new LS-P display may be easier for subjects to use, resulting in lower fatigue.

## 1 Introduction

Brain computer interfaces (BCIs) establish a communication pathway directly between human brains and external devices by recognizing voluntary changes in the users' brain activity, which can help severely disabled patients regain some functional independence and freedom to interact with people and their environments [1-5]. BCI systems have mostly relied on motor imagery [6, 7], steady state visual evoked potential (SSVEP) [8-12], slow cortical potentials (SCP) [13], and the P300 with other event-related potentials [14-17]. The P300-based BCI was first presented by Farwell and Donchin in 1988 [13], and has become one of the most prominent BCI approaches. In last two decades, many methods were explored to improve the P300-based BCI in terms of classification accuracy [18], information transfer rate [19], convenience, and comfort [20]. Optimized paradigms and algorithms were presented to enhance the difference between attended and ignored events, which could entail more recognizable differences in the P300 and/or other components and increase the classification accuracy and information transfer rate [21-22]. Some other paradigms were presented to meet the requirement of some special groups [23-25]. For example, gaze independent BCIs were designed for the patients whose eye movements were impaired [26].

Amyotrophic lateral sclerosis (ALS) is one of the most common motor neuron diseases. ALS patients progressively lose motor and speech functions due to declining motor control, eventually leaving them with few or no ways to move and thereby communicate [27]. BCIs have become a well-studied technique that could help these patients by providing communication that does not require any movement [28-33]. Sellers et al., 2006 tested P300-based BCIs with four orders on 3 ALS patients and reported that P300-based BCI can serve as a non-muscular communication device for ALS patients [34]. In 2010, Sellers and colleagues also studied extensive P300 speller use with ALS patients (6-8 hours a day for several months) [35]. Holz et al., 2015 also presented a brain painting BCI system for long term using by ALS patients [36]. Mak et al., 2012 surveyed the performance of a P300 BCI using a 36-character spelling system with ALS patients, and also showed that P300 spellers could work on ALS patients [37]. In that study, ALS subjects were required to spell 14 characters online in four runs. McCane et al., 2015 studied the differences between ALS patients and age matched healthy subjects when they used the P300-based BCI, and reported differences in classification accuracy and information transfer rate between these two groups [38]. Mainsah et al., 2015 tested P300-based BCIs on ALS patients using a dynamic stopping strategy. In this study, the ALS patients spelled 36 characters online in three sessions (12 characters/session) [39]. In 2015, a P300-based BCI that adopted the T9 speller approach [40] was tested on one locked-in ALS patient, who failed in the copy spelling of a common sentence due to a technical problem [41]. However, that ALS patient could spell a 23 character long sentence with only one mistake using T9 speller in 500 seconds. Face stimuli can yield higher performance with both healthy and patient subjects compared to flashing characters [20, 21]. Thus, there is ample work validating P300-based BCIs with ALS patients, although the extent to which these BCIs require control of eye gaze is controversial [42-44].

Gaze independent BCIs have been explored as an option for ALS patients who can no longer reliably control eye movements [44]. However, even gaze independent BCIs may not work on some of these patients [45, 46]. Oculomotor disorders from ALS include ophthalmoplegia, defective pursuit, saccadic impairments, nystagmus, and abnormal Bell phenomena, which vary with disease progression [47]. Persons with moderate ALS could have better outcomes with BCIs than persons with complete locked-in syndrome [46]. One practical question that has not been well explored is performance across continuous use. Patients might want to spell continuously over many minutes, and fatigue or other issues could reduce performance in a gaze-dependent BCI, particularly when patients must attend to stimuli that are several degrees from the fovea.

In this paper, we aimed to survey P300-based BCI performance when ALS patients participated in continuous spelling tasks across differing visual angles. We assessed continuous spelling with a paradigm using 42 characters per run. We surveyed the effects of the characters' visual angle by comparing two paradigms (the traditional matrix-based speller paradigm (MS-P) and a new large visual angle speller paradigm (LS-P)).

Table 1. Study participants' background information. "G" is Gender, "DC" is Diagnosis Categories, "CS" is Clinical stage, "ALSFRS-R" is ALS Functional Rating Scale, "m" is month, "B" is Bulbar, "C" is cervical, "T" is thoracic and "L" is lumbar.

| No  | G | Age       | CS  | Type of onset | ALSFRS-R | Duration(m) | EMG       |
|-----|---|-----------|-----|---------------|----------|-------------|-----------|
| S1  | M | 55        | 2,B | upper limb(C) | 34       | 45          | C,T(2)    |
| S2  | M | 69        | 2,B | upper limb(C) | 38       | 21          | C,T(2)    |
| S3  | M | 43        | 3   | lower limb(L) | 20       | 48          | B,L(2)    |
| S4  | F | 30        | 3   | upper limb(C) | 42       | 26          | B,C,T,L(4)|
| S5  | M | 42        | 2,A | upper limb(C) | 47       | 33          | B,C,T,L(4)|
| S6  | M | 48        | 3   | upper limb(C) | 19       | 56          | B,C,T,L(4)|
| S7  | F | 30        | 3   | upper limb(C) | 42       | 49          | B,C,T,L(4)|
| S8  | M | 59        | 3   | upper limb(C) | 27       | 20          | B,C,L(3)  |
| S9  | M | 72        | 2,B | upper limb(C) | 30       | 22          | B,C,T,L(4)|
| S10 | M | 70        | 2,B | upper limb(C) | 27       | 12          | B,C,T,L(4)|
| S11 | F | 62        | 2,A | lower limb(L) | 44       | 12          | T,L(2)    |
| S12 | M | 55        | 2,B | lower limb(L) | 46       | 12          | T,L(2)    |
| S13 | M | 69        | 2,B | upper limb(C) | 33       | 43          | B,C,L(3)  |
| S14 | M | 60        | 2,A | upper limb(C) | 37       | 36          | B,C,L(3)  |
| S15 | M | 58        | 2,B | bulbar        | 39       | 18          | B,C,L(3)  |
| S16 | M | 58        | 2,B | upper limb(C) | 40       | 11          | B,C,L(3)  |
| S17 | F | 66        | 3   | lower limb(L) | 25       | 10          | B,C,T,L(4)|
| S18 | M | 46        | 3   | upper limb(C) | 29       | 24          | B,C,T,L(4)|
| AVG |   | 55.1±12.9 |     |               | 34.4±8.6 | 27.7±15.1   |           |

## 2 Methods

*2.1 Subjects*

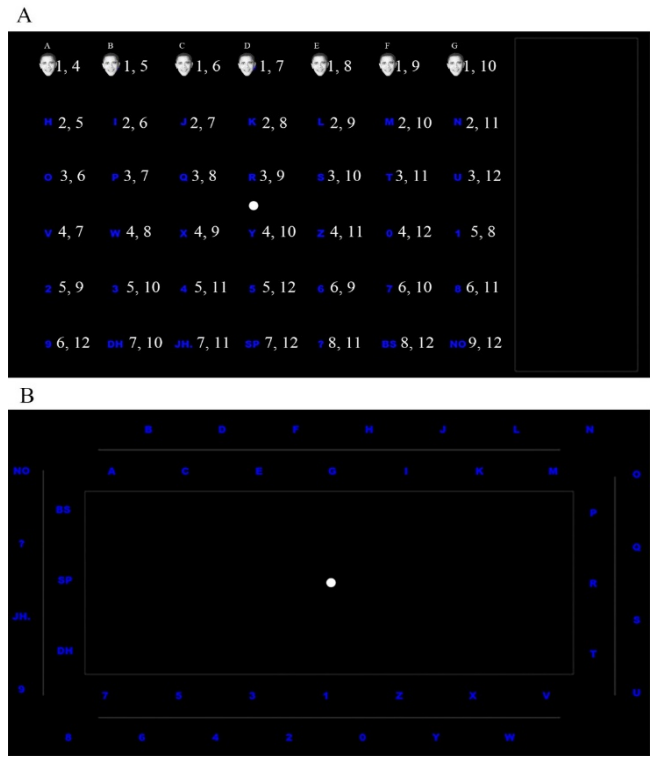

Fig.1 The displays used in the MS-P (panel A) and LS-P (panel B) conditions. The text and images in white are presented here to further explain the paradigm, and were not shown to the subjects

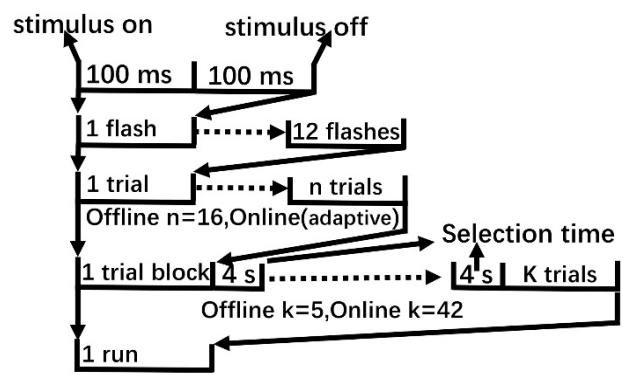

Fig.2 One run of the online and offline experiments

Eighteen ALS subjects (14 male, 4 female, aged 30 to 70 years, mean age 55.1±12.9) participated in this study (see Table 1). It was reported that about 30 % of patients with ALS exhibit cognitive impairments [48]. In our study, we communicated with all ALS patients before the experiment, and all subjects could understand our BCI task and perform the task correctly. Four stages are used to classify ALS patients [49]. ALS symptoms are mild in stage 1, whereas ALS patients in stage 4 A need gastrostomy and ALS patients in stage 4 B even require non-invasive ventilation. We selected ALS patients in stages 2 and 3. Only one of the subjects (S1) had experience with a BCI. Each subject participated in all sessions within one day. The order of the conditions was counterbalanced across subjects for each session. All subjects signed a written consent form prior to this experiment and were paid 150 RMB for their participation.

The local ethics committee approved the consent form and experimental procedure before any of the subjects participated. In the experiment, the subject was required to sit 80 cm in front of the display. We adjusted the display such that the center of each subject's field of view (0° visual angle) was in the center of stimulus matrix for both paradigms.

*2.2 Stimuli*

We use the term "flash groups" to refer to the method of grouping icons within each flash. There were 12 flash groups in total, and each flash group flashed a different subset of icons. Fig. 1 A shows the MS-P display presented to all subjects, which was a 6 × 7 matrix. The white numbers, circles and characters in Fig. 1 A and B are used to show when each corresponding stimulus flashed, and were not shown to the subjects. The cells in the matrix with same number would be arranged as one flash group. For example, fig 1.A shows that that the cells in the first flash group are flashing. All of the cells that flashed contain number "1".

The feedback was shown in the left side of the screen. Fig.1 B shows the display of LS-P presented to all subjects. The flash code of each character was same as the character in MS-P. The feedback was shown in the middle of the screen.

Sixteen subjects' data were recorded in the patients' homes or hospitals, while 2 subjects' data were recorded in our lab. Stimuli were presented to the patients using a laptop computer. The size of the display was 26 cm×19.5 cm. The target visual angle of MS-P was from 1.07° to 9.58° (6.0°±2.3°) and the target visual angel of LS-P was from 4.43° to 12.34° (8.8°±2.5°). DH" is ",", "JH." is ".", "SP" is "space", "BS" is "backspace", and "No" is "cancel".

*2.3 Experimental set up, offline and online protocols*

EEG signals were recorded with a g.USBamp and a g.EEGcap (Guger Technologies, Graz, Austria) with active electrodes, sampled at 256 Hz. The g.USBamp uses wide-range DC-coupled amplifier technology in combination with 24-bit sampling. The result is an input voltage of +/- 250 mV with a resolution of <60 nV. The impedance of the electrodes was less than 30 kΩ. Data were recorded and analyzed using the BCI platform software package developed through the East China University of Science and Technology. We recorded from 16 EEG electrode positions (F3, Fz, F4, FC1, FC2, C3, Cz, C4, P7, P3, Pz, P4, P8, O1, Oz, O2) based on the extended international 10-20 system. Active electrodes were referenced to the right ear, using a front electrode (FPz) as ground. The recorded data were filtered using a high pass of 0.5 Hz and a low pass of 30 Hz, notch-filtered at 50 Hz [39]. A prestimulus interval of 100 ms was used for baseline correction for single trials. Each subject participated in offline and then online runs of two conditions: MS-P and LS-P. The order of the conditions was counterbalanced. Both conditions used a "copy spelling task" during the offline and online testing described below, meaning that subjects were told which item was the target before each run.

The stimulus "on" time was 100 ms and the stimulus "off" time was 100 ms, yielding an SOA of 200 ms. As noted, each flash reflected each time a stimulus changed from its background to a face. One trial contained all flashes with each of the 12 flash patterns, and each trial lasted 2.4 s. A trial block referred to a group of trials with the same target. During offline testing, there were 16 trials per trial block and each run consisted of five trial blocks, each of which involved a different target (see fig. 2). The number of trials averaged in the online runs for each trial block was selected adaptively in each run [19]. The classifier must output the same result for two times to stop presenting additional trials.

Each subject first participated in three offline runs, with a 3 min break after each offline run. After all offline runs of the two conditions, subjects were tasked with attempting to communicate 42 targets (i.e. 42 trial blocks) without interruption for each condition in the online experiment (called one online run). Before each online run, we would ask each subject if he/she had enough rest and was ready for the online run. All the subjects were ready for the online run after a 3 minute break. Feedback and target selection time was 4 s before the beginning of each trial block.

*2.4 Feature extraction*

A third-order Butterworth band pass filter was used to filter the EEG between 1 and 30 Hz [50, 51]. The EEG was then down-sampled from 256 to 36.6 Hz by selecting every seventh sample from the filtered EEG. Single flashes lasting 800 ms were extracted from the data.

*2.6 Classification scheme*

Bayesian linear discriminant analysis (BLDA) is an extension of Fisher's linear discriminant analysis (FLDA) that avoids overfitting. BLDA was selected because of its demonstrated classification performance in P300 BCI applications [51] Data acquired offline were used to train the classifier using BLDA and obtain the classifier model. This model was then used in the online system.

*2.7 Statistical analyses*

We conducted paired-samples t-tests with the independent factor "paradigm" with two levels (MS-P and LS-P) and the dependent variable "class" for the ERP peaks, classification accuracy, information transfer rate, respectively. All the dependent variables were statistically tested for normal distribution (One-Sample Kolmogorov Smirnov test). The alpha level was $\alpha = 0.05$ (significant). In this paper, the online and offline accuracy, online bit rate, online trials for average, theta power and alpha power all meet normal distribution.

## 3 Result

We assessed two conditions with 18 ALS patients to survey the effects of visual angle and continuous use of an event related potential (ERP)-based BCI. These two conditions were called traditional matrix-based speller paradigm (MS-P) and large visual angle speller paradigm (LS-P). The mean visual angle of MS-P is 6.0° and the mean visual angle of LS-P is 8.8°. Fig. 3 shows the average ERP amplitudes elicited by the MS-P and LS-P conditions. Both two paradigms evoked clear N200, P300 and N400. Fig. 4 shows the offline classification accuracy of LS-P and MS-P. The result showed that there was no significant difference between LS-P and MS-P in classification accuracy (P>0.05).

Table 2 shows the online feedback accuracy, number of trials required to spell all 42 characters,

and bit rate for both MS-P and LS-P. "Feedback accuracy" was calculated based on the online output of the BCI system. There were no significant differences between the two paradigms in accuracy (p>0.05) and bit rate (p>0.05). There were 11 different visual angles in MS-P (two asymmetric targets (D and I) were located at same visual angle) and 12 different visual angles in LS-P, relative to the center of the display for each paradigm.

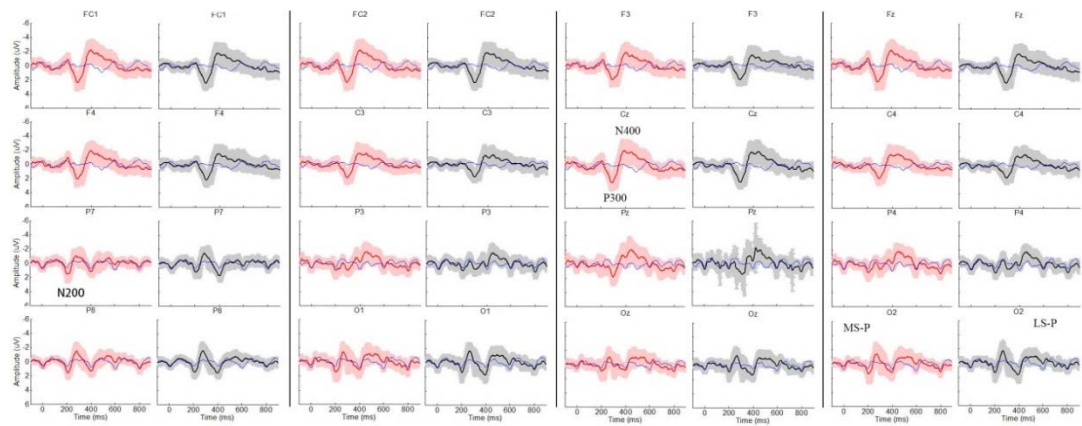

Fig.3 The average amplitude of event-related potentials for the MS-P and LS-P conditions. The red and gray dashed areas are the standard deviation.

Table 2. Online feedback accuracy number of trials used for spelling 42 characters and bit rate.

| Subject | Feedback Accuracy(%) | | Trials used for 42 targets | | Bit rate(bits/min) | |
|---|---|---|---|---|---|---|
| | MS-P | LS-P | MS-P | LS-P | MS-P | LS-P |
| S1 | 88.1 | 76.2 | 112 | 127 | 39.6 | 27.5 |
| S2 | 81.0 | 92.9 | 104 | 105 | 37.1 | 46.4 |
| S3 | 81.0 | 85.7 | 110 | 103 | 35.1 | 41.1 |
| S4 | 59.5 | 88.1 | 121 | 120 | 19.5 | 37.0 |
| S5 | 61.9 | 71.4 | 131 | 136 | 19.2 | 23.1 |
| S6 | 64.3 | 47.6 | 133 | 138 | 20.0 | 12.1 |
| S7 | 78.6 | 83.3 | 110 | 119 | 33.4 | 33.9 |
| S8 | 100.0 | 97.6 | 94 | 91 | 60.2 | 58.9 |
| S9 | 90.5 | 97.6 | 95 | 95 | 49.0 | 56.4 |
| S10 | 85.7 | 100.0 | 104 | 99 | 40.7 | 57.2 |
| S11 | 81.0 | 69.0 | 107 | 121 | 36.0 | 24.6 |
| S12 | 90.5 | 92.9 | 95 | 102 | 49.0 | 47.8 |
| S13 | 92.9 | 81.0 | 100 | 112 | 48.7 | 34.4 |
| S14 | 83.2 | 92.9 | 98 | 99 | 41.1 | 49.2 |
| S15 | 81.0 | 85.7 | 125 | 119 | 30.9 | 35.6 |
| S16 | 95.2 | 100 | 107 | 91 | 47.7 | 62.2 |
| S17 | 71.4 | 61.9 | 115 | 128 | 27.4 | 19.6 |
| S18 | 90.5 | 78.6 | 105 | 103 | 44.3 | 35.6 |
| AVG | 82.0±11.5 | 83.5±14.2 | 109.2±11.9 | 111.6±14.9 | 37.7±11.4 | 39.0±12.1 |
| | P>0.05 | | P>0.05 | | P>0.05 | |

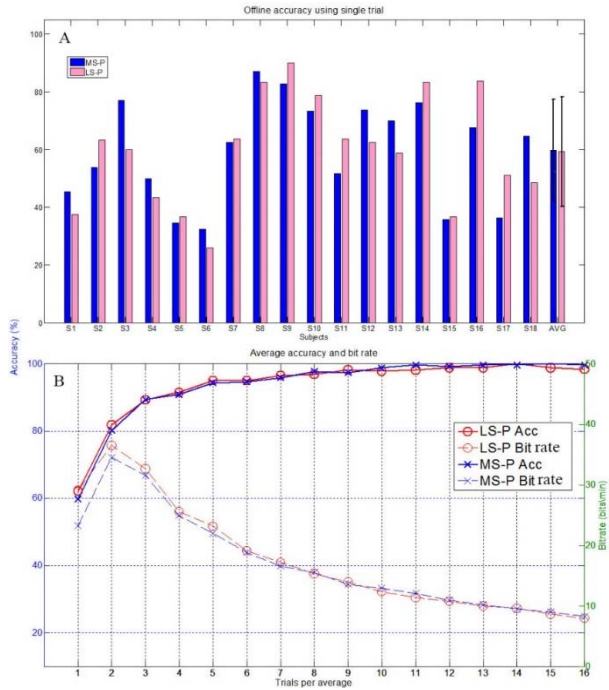

Fig.4 Offline classification accuracy. Panel A is the classification accuracy for each subject based on single trials, and panel B is the average classification accuracy for all subjects across averages of 1-16 trials.

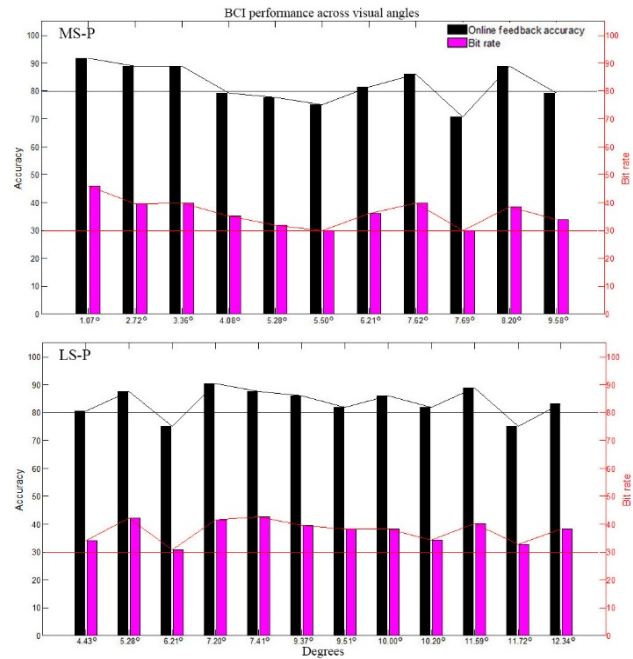

Fig.5 Online feedback accuracy and bit rate for different visual angles for both paradigms, MS-P and LS-P.

Fig. 5 shows the online feedback accuracy and bit rate across different visual angles. Fig. 5 indicates that changes in visual angle did not affect BCI accuracy or bit rate in our study. Since the number of samples was less than 30, we used bivariate correlation analyses with Spearman coefficients to explore the correlation between online feedback accuracy and visual angle. There were no significant correlations between online feedback accuracy and visual angle (r=-

0.446, p>0.05 for MS-P, r=-0.067, p>0.05 for LS-P), and between bit rate and visual angle (r=-0.455, p>0.05 for MS-P, r=-0.119, p>0.05).

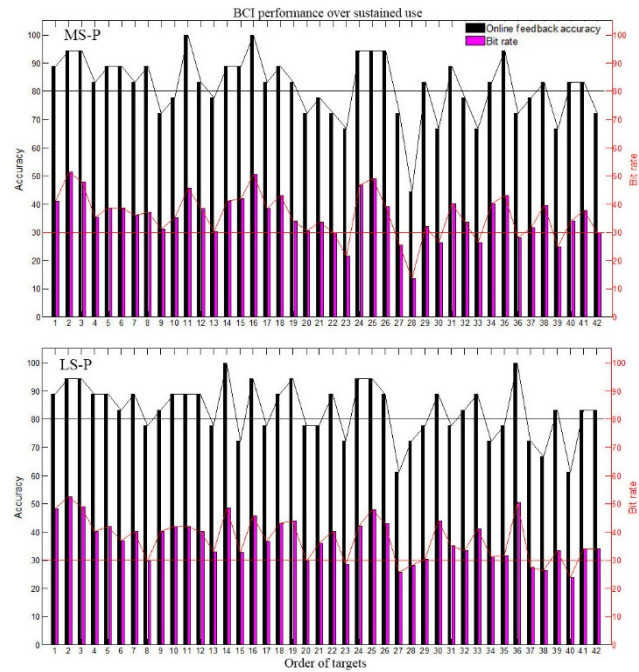

Fig.6 Online feedback accuracy and bit rate for 42 targets in chronological order.

Online BCI performance did decline slightly with continuous use in this study. Fig. 6 shows the online feedback accuracy and bit rate across the 42 targets in the chronological order they were presented to subjects. 4 characters were classified below 80% in the first 21 characters (86.0±7.8%), and 11 characters were classified below 80% in last 21 characters (78.0±12.5%) for MS-P. Six characters were classified below 80% in the first 21 characters (86.5±7.4%), whereas ten characters were classified below 80% in the last characters (80.4±10.8%) for LS-P. In the MS-P paradigm, none of the first 21 characters were below 30 bits/min (39.1±6.1 bits/min), while 9 characters were below 30 bits/min in the last 21 characters (33.0±8.7 bits/min). In the LS-P paradigm, 2 of the first 21 characters were below 30 bits/min (40.6±6.3 bits/min), and 6 of the last 21 fell below 30 bits/min (34.8±7.6 bits/min).

The theta power at Fz and alpha power at Pz [52] from the first 21 targets vs. the last 21 targets were calculated for MS-P and LS-P (see fig. 7). The result showed a significant increase of power in the alpha band ($p<0.01$) in MS-P, and no significant difference in alpha power was found in LS-P, which suggests that the LS-P condition entailed lower fatigue than MS-P.

We conducted bivariate correlation analyses using Pearson coefficients between both measures of online performance (accuracy and bit rate) and character order (the order in which the characters were presented to the patients). There were significant negative correlations between accuracy and character order (r=-0.373, p<0.05 for MS-P and r=-0.395, p<0.01 for LS-P), and between bit rate and character order (r=-0.385, p<0.05 for MS-P and r=-0.499, p<0.01 for LS-P).

## 4   Discussion

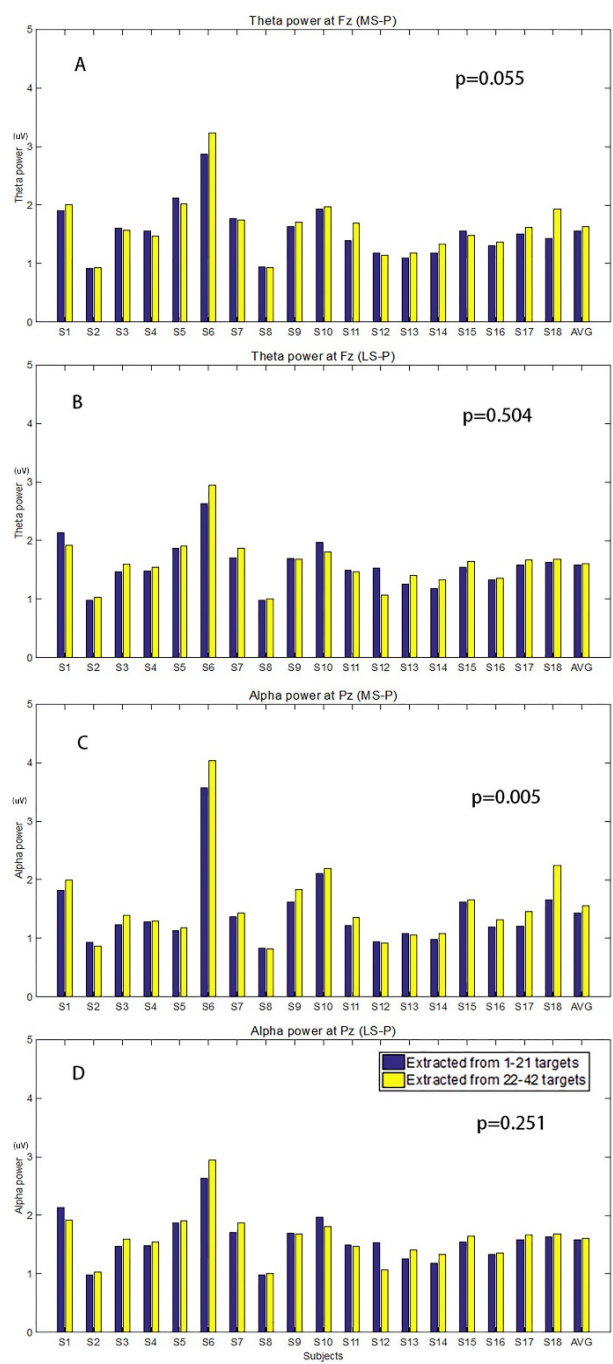

Fig.7 The theta and alpha power resulting from the first 21 targets (blue) and the last 21 targets (yellow) MS-P and LS-P.

While early BCI work with ALS patients did not use ERP-based BCIs [52, 53], ERP-based BCIs have since become more prominent tools for ALS patients, due to their high classification accuracy and information transfer rate, reliability with most users, and very low training requirements. 56% of BCI studies on ALS patients used ERP-based BCIs, as of 2015 [55]. This work has shown that ERP-based BCIs can be useful tools for ALS patients, but has also exposed several unanswered questions.

In this paper, two ERP-based BCI systems were tested on 18 ALS patients during continuous use. Comparable results were obtained with a conventional paradigm (MS-P) and a novel paradigm using a larger visual angle (LS-P) with ALS patients (see Table 2).

*4.1 Visual angle effect*

Two ERP-based BCI paradigms with different visual angles were tested to assess the impact of higher visual angles, which might influence gaze dependence and fatigue. The mean visual angle of LS-P was 2° larger than that of MS-P. Table 2 showed that there were no significant differences (all p>0.05) between these two paradigms in feedback accuracy or bit rate. Figs. 4 through 6 concordantly show no significant differences in offline classification accuracy, online classification accuracy, and bit rate, as well as no clear feedback accuracy and bit rate trends that correspond to any difference between the paradigms. These results indicate that the visual angle did not affect the performance of this ERP-based BCI for these ALS patients when the visual angle of the stimulus was within 12.34° of the center, which was the largest visual angle used in our study.

*4.2 Continuous use of BCI*

We were not aware of work that assessed the performance of ERP-based BCI system when more than 30 characters were tested without any pause. In this study, ALS patients could spell 42 characters online, requiring more than 7 minutes. Fig. 6 and the associated correlation analyses show that accuracy and bit rate did decline slightly during the second half of the run. However, 50% of subjects could obtain more than 90% classification accuracy in one of the paradigms (33.3% for MS-P and 38.9% for LS-P) and 88.3% subjects could obtain more than 80% classification accuracy in one of the paradigm (72.2% for MS-P and 66.6% for LS-P). Our results with continuous use indicate that spelling for about seven minutes is feasible, but a short break mid-run could improve performance. Thus, individual users might customize the duration of each run accordingly. Käthner et al.2010 studied the effects from fatigue across 10 blocks, with breaks between blocks, and compared the pre and post state. They reported that longer spelling sessions lead to lower ERP amplitude and BCI performance [52]. In our study, we obtained a similar result by analyzing the classification trend in one 7 minutes block with ALS patients. It was found that the alpha power increased significantly (p<0.01) in MS-P and alpha power was relatively stable in LS-P (p>0.05), which indicated that the MS-P condition entailed higher fatigue than LS-P. The reasons may involve display size and/or some other differences in the display layout (such as the blank feedback area in the center of LS-P only). Now that this initial effect has been identified, we hope to further assess the underlying causes.

*4.3 Patient variability*

The patients in this study had some residual motor control. This may raise the question of whether this study's results are relevant, since these patients might be able to use other assistive technologies (ATs) to communicate instead of a BCI. However, ALS patients in stages 2 and especially 3 may find speech, eye movements, or other muscle activity fatiguing, and might switch to a BCI when they are tired [56-58]. Other AT systems may be impractical for various reasons; notably, several of our subjects did not consistently use AT to communicate. Thus, even for ALS patients like those we studied here, BCIs could be helpful, especially if BCIs continue to improve in terms of portability, cost, speed, accuracy, flexibility, and ease of use.

Prior work has suggested that our results may be different with patients with more severe disabilities. Two meta-analyses found that completely locked-in state (CLIS) patients could not effectively use BCIs [46][55]. [46] found that, once the CLIS patients were removed from analysis, the remaining patients did not exhibit any correlation between BCI performance and physical impairment. We agree with their conclusion that training patients to use BCIs while they have some remaining ability to learn BCI use is very important. Training could also help patients improve covert attention and thereby reduce dependence on gaze, though this has not been explored. This early BCI use also provides EEG data that could be used to improve patient-specific classifiers [59] and perhaps lead to other benefits, such as improved diagnosis and treatment or new BCI options. In addition to motor disabilities, cognitive disabilities [48] and low motivation could also affect BCI performance [60][61].

*4.4 Use of large visual angle speller paradigm*
Piccione et al., 2006 firstly used the P300 BCI with large visual angle speller paradigm on paralysed patients, when the control object was in the middle of the screen [62]. Bai et al., 2015 also used the large visual angle speller paradigm to control the explorer [63]. Large visual angle speller paradigm with one screen would be more comfortable when the controller object was in the middle of the screen compared to two screens, when one was used for BCI system and another was used for control object [64]. Fig. 7 also indicated that the new LS-P display may be easier for subjects to use, resulting in lower fatigue.

5  Conclusion

Two ERP-based BCI systems were tested on ALS patients with moderate disabilities to explore the effects of continuous use and visual angle. Increasing visual angle did not significantly affect BCI performance. 88.3% of subjects still could obtain more than 80% classification accuracy during successive BCI use, but performance declined slightly over time. In future work, we will further assess successive use as well as users' needs and preferences; some users may not want or like long sessions. We will also test our system on ALS patients in stage 4.

6. Acknowledgment

This work was supported in part by the Grant National Natural Science Foundation of China, under Grant Nos. 91420302, 61573142. This work was also supported by the Fundamental Research Funds for the Central Universities (WH1516018) and Shanghai Chenguang Program under Grant 14CG31.